\documentclass{jpp}
\usepackage{graphicx,color,amsmath,amsfonts,amssymb,mathtools}
\usepackage{epstopdf,subfigure, epsfig,mathrsfs,psfrag,mathrsfs}
\usepackage{wasysym}
\usepackage{marvosym}
\usepackage{tikz}

\newtheorem{theorem}{Theorem}

\newcommand{\be}{\begin{equation}}
\newcommand{\en}{\end{equation}}

\def\brho{{\boldsymbol\rho}}
\def\d{{\rm d}}
\def\uv{{\boldsymbol u}}
\def\tv{{\boldsymbol t}}

\def\xv{{\boldsymbol x}}
\def\yv{{\boldsymbol y}}
\def\nv{{\boldsymbol n}}

\def\wv{{\boldsymbol w}}

\def\Bv{{\boldsymbol B}}

\def\Pv{{\boldsymbol P}}
\def\Av{{\boldsymbol A}}

\def\grad{\boldsymbol\grad}

\def\wh{\mathcal{H}}

\def\bs{{\rm BS}}

\def\div{{\rm div}\, }
\def\grad{{\rm grad}\, }
\def\curl{{\rm curl}\, }

\shorttitle{Helicity in multiply connected domains}
\shortauthor{D. MacTaggart and A. Valli}

\title{Magnetic helicity in multiply connected domains}

\author{D. MacTaggart\aff{1}\aff{2}
\corresp{\email{david.mactaggart@glasgow.ac.uk}}
\and A. Valli\aff{1}
}

\affiliation{\aff{1}Department of Mathematics, University of Trento, Povo, Italy
\aff{2}School of Mathematics and Statistics, University of Glasgow, Glasgow, G12 8QW, UK}

\begin{document}

\maketitle

\begin{abstract}
Magnetic helicity is a fundamental quantity of magnetohydrodynamics that carries topological information about the magnetic field. By `topological information', we usually refer to the linkage of magnetic field lines. For domains that are not simply connected, however, helicity also depends on the topology of the domain. In this paper, we expand the standard definition of magnetic helicity in simply connected domains to multiply connected domains in $\mathbb{R}^3$ of arbitrary topology. We also discuss how using the classic Biot-Savart operator simplifies the expression for helicity and how domain topology affects the physical interpretation of helicity.
\end{abstract}

\begin{keywords}
Magnetic helicity, Multiply connected domains, Magnetic topology
\end{keywords}

\section{Introduction}
\label{intro}
The classical definition of magnetic helicity in magnetohydrodynamics (MHD) is for a bounded \emph{simply connected} domain. Let $\Omega$ be such a domain with boundary $\partial\Omega$ and outward unit normal $\nv$. Consider a magnetic field $\Bv$ in the space
\be\label{vspace}
\mathcal{V} = \{\Bv\in(L^2(\Omega))^3: \div\Bv=0, \ \Bv\cdot\nv=0\}.
\en
Then the magnetic helicity (hereafter helicity) is
\be\label{hel_sc}
H = \int_{\Omega}\Av\cdot\Bv,
\en
where $\Bv=\curl\Av$. It is well-known that $H$ is an invariant of ideal MHD \citep{woltjer58} and is also well-conserved in resistive MHD with very small magnetic diffusion \citep{berger84,faraco19}. It is also well-known that $H$ is \emph{gauge invariant}, i.e. $H$ is not affected by the change $\Av\rightarrow\Av+\grad\chi$, for some scalar function $\chi$.

The topological interpretation of $H$ is due, originally, to \cite{moffatt69}. By considering the Coloumb gauge, $\div\Av=0$, \cite{moffatt69} showed that $H$ could be interpreted in terms of the Gauss linking number weighted by magnetic flux. The linking number can refer to the linkage of thin flux tubes \citep[e.g.][]{moffatt69} or field lines \citep[e.g.][]{arnold92}. Since both helicity and magnetic flux are invariants of ideal MHD, this linkage is also invariant. The topology of field lines has also been studied in magnetic fields that are not everywhere tangent to the domain boundary \citep[e.g.][]{bergerfield84} but we will not consider such fields in this work. 

Moving to \emph{multiply connected} domains, there have been two general approaches to determining helicity. The first comes from plasma physics and focusses on a toroidal domain suitable for fusion devices \citep[e.g.][]{taylor15}. This domain is not simply connected and $H$ from equation (\ref{hel_sc}) is not (in general) gauge invariant in such a domain.  The main reference in the plasma physics literature to the gauge invariant form of helicity in a torus is \cite{bevir80}, who state that the correct form of helicity is
\be\label{hel_tor}
H=\int_{\Omega}\Av\cdot\Bv -\oint_{\gamma_1}\Av\cdot\tv_1\oint_{\gamma_2}\Av\cdot\tv_2,
\en
where $\Omega$ no longer refers to a simply connected domain (which will be assumed for the rest of this work), $\gamma_1$ and $\gamma_2$ are closed paths on $\partial\Omega$ travelling around the major and minor circumferences of the torus, respectively, and $\tv_1$ and $\tv_2$ are the associated {unit} tangent vectors of the paths. The derivation of equation (\ref{hel_tor}), which we refer to as the \emph{Bevir-Gray formula}, in the plasma physics literature is based on making the transformation $\Av\rightarrow\Av+\grad\chi$ but with $\chi$ now being multivalued \citep[e.g.][]{biskamp93,marsh96}. After the transformation is performed, the domain is cut in order to make $\chi$ single valued and the result is the second term on the right-hand side of equation (\ref{hel_tor}). This approach is difficult to generalize to domains of more complex topology. Later, we will derive a generalized version of the Bevir-Gray formula by means of the Helmholtz decomposition of vectors in multiply connected domains.

Whereas the approach in plasma physics has been to focus on a particular domain and not specify a particular vector potential, the second approach to determining helicity in multiply connected domains has been to specify the vector potential and consider arbitrary domains. In a series of papers by Cantarella and collaborators \citep{cantarella00,cantarella00b,cantarella01,cantarella02}, the vector potential is chosen to satisfy the Coloumb gauge and takes the form of the \emph{Biot-Savart operator},
\be\label{bs_op}
\bs(\Bv)(\xv) = \frac{1}{4\pi}\int_{\Omega}\Bv(\yv)\times\frac{\xv-\yv}{|\xv-\yv|^3}\,\d\yv.
\en
This operator appears in other contexts in electromagnetism and, as mentioned previously, was used in \cite{moffatt69} to interpret helicity in terms of the Gauss linking number.\footnote{When Gauss' formula was published officially in 1867, it was included as part of a collection of material relating to electromagnetic induction \citep{epple98}. This represents one of the many early links between topology and electromagnetism.} One particularly interesting property of the Biot-Savart operator is that it can be made self-adjoint, leading to the application of spectral theory where magnetic fields that maximize the helicity in a domain correspond to linear force-free fields \citep[e.g.][]{cantarella00,valli19}. In this paper, we will refer to helicity expressed in terms of the Biot-Savart operator as \emph{Biot-Savart helicity}. This term is purely for convenience and and does not imply a new object, i.e. Biot-Savart helicity is still helicity but written in a particular way.


In this work we show that the two approaches to finding helicity in multiply connected domains, described above, can be unified. We first generalize equation (\ref{hel_tor}) to connected domains of arbitrary topology in a systematic way. We then show how the gauge invariant helicity in multiply connected domains is coincident with Biot-Savart helicity. {This leads to a discussion of how the, often quoted, \emph{mutual helicity} \citep{moffatt69,laurence93,cantarella00b} follows from the general helicity formula.}  The paper ends with a summary and a discussion of the interpretation of helicity in periodic domains.

\section{Helicity in multiply connected domains}
\subsection{Geometrical setup}\label{sec:geo_setup}
We now describe the general geometical setup and introduce ideas from homology which are necessary for treating multiply connected domains. A comprehensive review of the application of homology to magnetic fields can be found in \cite{blank57}. A more recent and accessible account can be found in \cite{cantarella02}.

We consider a domain $\Omega\in\mathbb{R}^3$ that is a bounded open connected set with Lipschitz continuous boundary $\partial\Omega$ and outer unit normal vector $\nv$. If the first Betti number of $\Omega$ is (the genus) $g>0$, then the first Betti number of $\partial\Omega$ is $2g$ \citep[e.g.][]{cantarella02}. We can consider $2g$ non-bounding cycles on $\partial\Omega$, $\{\gamma_j\}^g_{j=1}\cup\{\gamma'_j\}^g_{j=1}$, that represent the generators of the first homology group of $\partial\Omega$. Each set of cycles is associated with the closed domain and its complement in a ball containing the domain. $\{\gamma_j\}^g_{j=1}$ represent the generators of the first homology group of $\overline{\Omega}$ and have {unit} tangent vectors denoted by $\tv_j$. $\{\gamma'_j\}^g_{j=1}$ represent the generators of the first homology group of {$\overline{\Omega'}$, where $\Omega'=B\setminus \overline{\Omega}$ and $B$} is an open ball containing $\overline{\Omega}$. The {unit} tangent vector of a cycle $\gamma'_j$ is denoted $\tv'_j$. 

\begin{figure}
	\centering
	\begin{tikzpicture}[scale=1.2]  
\begin{scope}[scale=0.8]
\path[rounded corners=24pt] (-.9,0)--(0,.6)--(.9,0) (-.9,0)--(0,-.56)--    (.9,0);
\draw[rounded corners=28pt] (-1.1,.1)--(0,-.6)--(1.1,.1);
\draw[rounded corners=24pt] (-.9,0)--(0,.6)--(.9,0);


\draw[cyan] (0,0) ellipse (1.3cm and 0.55cm);
\draw[cyan] (3.8,0) ellipse (1.3cm and 0.55cm);

\end{scope}
\draw[densely dashed,brown] (0.19,-0.79) arc (270:90:.2 and 0.29);
\draw[brown] (0.19,-0.79) arc (-90:90:.2 and .290);
\draw(1.0605,0.5656) arc (45:315:1.5 and 0.8);
\draw(1.0605,0.5656) to[out=-28.1,in=200] (2,.6);
\draw(1.0605,-0.5656) to[out=28.1,in=160] (2,-.5);
\draw(2,-0.5) arc (45:315:-1.5 and -0.779);
\draw[densely dashed,brown] (3,0.25) arc (270:90:.2 and 0.29);
\draw[brown] (3,0.25) arc (-90:90:.2 and .290);

\begin{scope}[scale=0.8]
\draw[rounded corners=28pt] (3.8-1.1,.1)--(3.8+0,-.6)--(3.8+1.1,.1);
\draw[rounded corners=24pt] (3.8-.9,0)--(3.8+0,.6)--(3.8+.9,0);
\end{scope}

\node at (-0.4,0.6) {$\textcolor{cyan}{\gamma_1}$};
\node at (0.4,-1) {$\textcolor{brown}{\gamma'_1}$};
\node at (3.25,1) {$\textcolor{brown}{\gamma'_2}$};
\node at (3.75,0.52) {$\textcolor{cyan}{\gamma_2}$};

\end{tikzpicture}

\caption{A domain with $g=2$. The cycles are shown in brown and cyan, following the notation in the main text. The oriented surfaces bounded by {the cycles  $\gamma_j'$ and $ \gamma_j$ define the cutting surfaces $\Sigma_j$ and $\Sigma'_j$ respectively.}}
\label{geo_setup}
\end{figure}
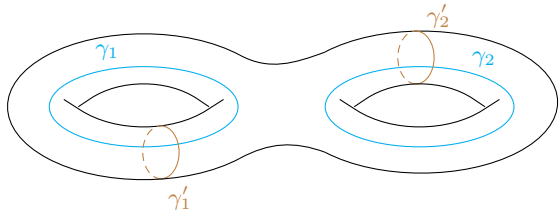

In $\Omega$ there are $g$ cutting surfaces, $\{\Sigma\}^g_{j=1}$, that are connected orientable Lipschitz surfaces satisfying $\Sigma_j\subset\Omega$ and $\partial\Sigma_j\subset\partial\Omega$.  Each surface $\Sigma_j$ satisfies $\partial\Sigma_j=\gamma'_j$ {and cuts the cycle $\gamma_j$}. A similar set of cutting surfaces, $\{\Sigma'\}^g_{j=1}$, exists in $\Omega'$. An illustration of the geometrical setup for a domain with $g=2$ is shown in Figure \ref{geo_setup}. { For the sake of definiteness, we choose the unit normal vector $\nv_{\Sigma_j}$ on $\Sigma_j$ oriented in such a way that (1) the unit tangent vector $\tv_j'$ on $\gamma'_j=\partial\Sigma_j$ is oriented counterclockwise with respect to $\nv_{\Sigma_j}$ (the `right-hand rule') and (2) the unit tangent vector $\tv_j$ crosses $\Sigma_j$ consistently with the direction of $\nv_{\Sigma_j}$.}

\begin{figure}
	\centering
		\subfigure[{Toroidal shell with cutting surfaces}\label{shell}]{
	\begin{tikzpicture}[scale=1.5]
\begin{scope}[scale=0.8]
\draw(-2,-1) circle (1cm);
\draw(2,-1) circle (1cm);
\coordinate (c1) at (-1,-1,0);
\coordinate(c2) at (1,-1,0);
\draw (c1) to [bend left=80] (c2);
\draw[dashed] (c1) to [bend right=80] (c2);
\coordinate (c3) at (-3,-1,0);
\coordinate (c4) at (3,-1,0);
\draw (c3) to [bend left=90] (c4);

\draw(-2,-1) circle (0.5cm);
\draw(2,-1) circle (0.5cm);
\coordinate (c5) at (-2.5,-1);
\coordinate (c6) at (-1.5,-0.5);

\draw[dashed] (-1.5,-1) to [bend right=80] (1.5,-1);
\node at (0.0,-1.73) {$\Sigma_1$};
\node at (-2.75,-1) {$\Sigma_2$};
\draw (-2.5,-1) arc (180:120:1.15);
\draw (2.5,-1) arc (0:60:1.15);
\draw (1.5,-1) arc (0:60:0.66);
\draw (-1.5,-1) arc (180:120:0.66);
\end{scope}
\end{tikzpicture}}

\subfigure[{Major cross section}\label{major}]{
\begin{tikzpicture}[scale=1.5]
\begin{scope}[scale=0.8]
\draw(0,0) circle (2cm);
\draw(0,0) circle (0.5cm);
\draw[red,thick](0,0) circle (0.55cm);
\draw[densely dashed](0,0) circle (1.4cm);
\draw[densely dashed](0,0) circle (1.6cm);
\draw[blue,thick](0,0) circle (1.35cm);

\node at (0.1,0.9) {$\gamma'_1$};
\node at (0.0,-0.9) {$\Sigma_1$};
\draw[->](0.2,0.9) -- (0.6,1.2);
\draw[->](0.1,0.8) -- (0.1,0.54);
\end{scope}
\end{tikzpicture}}
\hspace{0.5cm}
\subfigure[{Minor cross section}\label{minor}]{
\begin{tikzpicture}[scale=1.5]
\begin{scope}[scale=0.8]
\draw(0,0) circle (2cm);
\draw(0,0) circle (0.5cm);
\draw[yellow!40!red,thick](0,0) circle (0.55cm);
\draw[green!80!black,thick](0,0) circle (1.95cm);
\node at (0.1,0.9) {$\gamma'_2$};
\node at (0.0,-1.1) {$\Sigma_2$};
\draw[->](0.2,0.9) -- (1.0,1.6);
\draw[->](0.1,0.8) -- (0.1,0.54);
\end{scope}
\end{tikzpicture}}
\caption{Toroidal shell with cutting surfaces. (a) A three dimensional illustration of a toroidal shell cut in half. The cutting surfaces $\Sigma_1$ and $\Sigma_2$ are indicated. (b) The major cross section (toroidal hole shown as dashed lines) where $\gamma'_1$ is the boundary of the annulus $\Sigma_1$ represented by the blue and red cycles. (c) The minor cross section where $\gamma_2'$ is the boundary of the annulus $\Sigma_2$ represented by the orange and green cycles. Note that the cross sections are not to scale.}   \label{hom}
\end{figure}
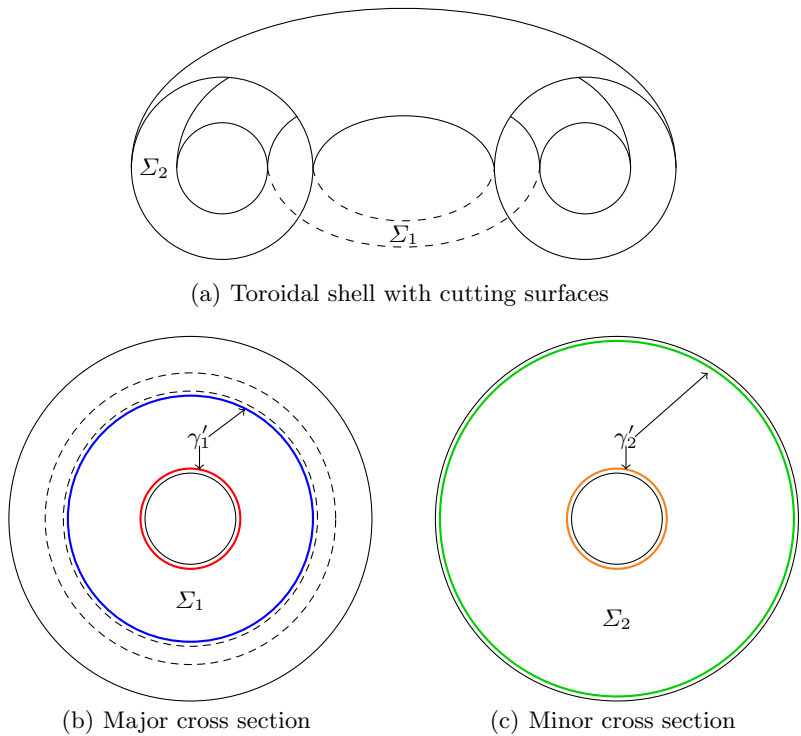

In the above description, if the boundary of $\Omega$ is \emph{not connected} then some care needs to be taken to make sure that a particular cutting surface remains in $\Omega$ or $\Omega'$ but not in both. For the $n$-holed torus, such as the 2-holed torus in Figure \ref{geo_setup}, this is not an issue. For a toroidal shell \citep[e.g.][]{oneil18}, however, the boundary is not connected, i.e. the outer boundary of the solid torus is separated from the inner boundary of the toroidal hole. A toroidal shell has $g=2$ and so there are two cycles for $\Omega$ and two for $\Omega'$. It is easy to see that the cutting surfaces $\Sigma'_j$, corresponding to the $\gamma_j$, lie entirely in $\Omega'$. Naively performing the same procedure for the cutting surfaces $\Sigma_j$ of the $\gamma_j'$, however, results in their overlapping with parts of the complementary domain $\Omega'$. Restricting the $\Sigma_j$ to lie entirely in $\Omega$ can be achieved using homological properties. We explain the procedure by describing the cutting surface for the cycles, $\gamma'_1$ and $\gamma_2'$, that orbit the central hole of the torus and the toroidal hole respectively. This situation is illustrated in Figure \ref{hom}.

The boundary of the first cutting surface, $\partial\Sigma_1=\gamma'_1$, is indicated in Figure \ref{major} by the red and blue cycles. Similarly, the boundary of the other cutting surface, $\partial\Sigma_2=\gamma'_2$, is indicated in Figure \ref{minor} by orange and green cycles.   The surfaces $\Sigma_1$ and $\Sigma_2$ lie entirely in $\Omega$ and are annuli. 

For a given orientation of the normal $\nv_1$ of $\Sigma_1$,
\be
\int_{\Sigma_1}\curl\wv\cdot\nv_1 = \int_{\rm blue}\wv\cdot\tv_1' - \int_{\rm red}\wv\cdot\tv_1',
\en 
where $\wv$ is a vector field and `blue' and `red' represent the cycles displayed in Figure \ref{major}. A similar result holds for $\Sigma_2$. Note, however, that the red cycle is bounding in $\Omega'$, thus $\gamma_1'$ and the blue cycle are equivalent homologically. The same is true for $\gamma_2'$ and the green cycle in Figure \ref{minor}.

\begin{figure}

	\centering
	
		{\includegraphics[scale=0.7,keepaspectratio]{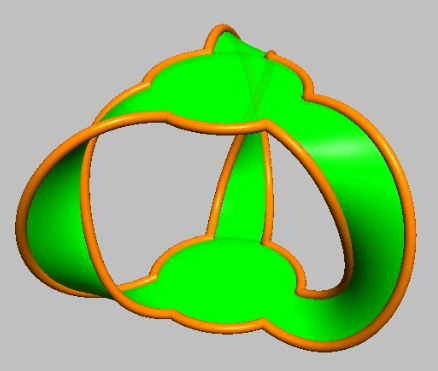}}
	\caption{An example of a `cut' domain that is not simply connected. Following the description in the main text, the orange domain is the trefoil knot $K$. The green surface is one of the cutting surfaces, shown here as two `discs' with three `twisting bands'. This image was produced with SeifertView (Jarke J. van Wijk, Technische Universiteit Eindhoven).}
\label{trefoil}
\end{figure}

In this paper, we will focus on ($n$-holed) tori and toroidal shells since these domains have the most immediate applications, {both as domains in their own right \citep[e.g.][]{dewar15,oneil18} and in their connection to the common simulation domains of periodic and doubly-periodic cubes (we will return to discuss helicity in periodic domains later)}. The `cut' domains (those formed by removing the cutting surfaces) for these cases are simply connected. This fact, however, is not true of all domains in $\mathbb{R}^3$. For example, consider a  domain $\Omega = B\setminus K$ where $B$ is a ball, as before, and $K$ is a trefoil knot \citep{benedetti12,alonso18}. {An illustration of such a domain is displayed in Figure \ref{trefoil}.} The procedure that we will now describe can cope with all the cases described above, including this last one.


\subsection{Helmholtz decomposition and Neumann harmonic vector fields}
The gauge transformation normally used in helicity studies, i.e. $\Av\rightarrow\Av+\grad\chi$, is only applicable in simply connected domains if $\chi$ is to remain single valued. For the regions under consideration with more complex topologies, we require the full Helmholtz decomposition \citep[e.g.][]{blank57,cantarella02}.

\begin{theorem}[Helmholtz]\label{helmholtz}
Any $\uv\in(L^2(\Omega))^3$ can be decomposed as
\be
\uv = \curl\Pv + \grad\phi+\brho,
\en 

where $\Pv$ is a vector field, $\phi$ a single valued scalar function and $\brho$ is in the space of Neumann harmonic fields $\wh$, defined as
\be
\wh = \{\brho\in(L^2(\Omega))^3:\curl\brho=\boldsymbol{0},\ \div\brho=0, \ \brho\cdot\nv=0\}.
\en
\end{theorem} 
In order to make use of $\wh$, we need its basis $\{\brho_j\}_{j=1}^g$. This can be found, for both $\Omega$ and $\Omega'$ separately, by {finding the solutions $\phi_j$ of} suitable elliptic problems where the appropriate cutting surfaces are removed from the domain under study. The basis functions for the space of harmonic fields in $\Omega$ take the form $\brho_j= \widetilde{\grad}\phi_j$, where  $\widetilde{\grad}\phi_j$ is the extension of ${\grad}\phi_j$ to $(L^2(\Omega))^3$ {and each $\phi_j$ has a jump equal to 1 on the corresponding cutting surface}. The basis functions of the space of harmonic fields in $\Omega'$ are constructed in a similar way and are denoted by  $\{\brho'_j\}_{j=1}^g$. For more details on the contruction of the basis functions, we direct the reader to \cite{alonso18}.
%

\subsection{Zero flux magnetic fields}
Consider two different vector potentials $\Av_1$ and $\Av_2$ for $\Bv\in\mathcal{V}$. Since $\curl(\Av_1-\Av_2)=\boldsymbol{0}$ in $\Omega$, it follows from Theorem \ref{helmholtz} that we can write
\be
\Av_1-\Av_2 = \grad\chi +\brho \ {\rm in} \ \Omega,
\en
where $\chi$ is a scalar function and $\brho\in\wh$. Some simple manipulation reveals
\be
\int_{\Omega}\Av_1\cdot\Bv-\int_{\Omega}\Av_2\cdot\Bv = \int_{\Omega}\Bv\cdot\brho.
\en
Therefore, if $\Bv\perp\wh$ then helicity is independent of the vector potential. Writing $\brho$ in terms of its basis functions, we require that
\be
\int_{\Omega}\Bv\cdot\brho_j = 0, \ {\rm for}\ j=1,\dots,g.
\en
This statement is nothing more than enforcing zero magnetic flux through every cutting surface,
\begin{eqnarray}
\int_{\Omega}\Bv\cdot\brho_j &=& \int_{\Omega\setminus\Sigma_j}\Bv\cdot\brho_j = \int_{\Omega\setminus\Sigma_j}\Bv\cdot\grad\phi_j \nonumber \\
&=& \int_{\partial(\Omega\setminus\Sigma_j)}\Bv\cdot\nv_j\phi_j -\int_{\Omega\setminus\Sigma_j}(\div\Bv)\phi_j \nonumber \\
 &=& \int_{\Sigma_j}\Bv\cdot\nv_j{[[}\phi_j{]]} = \int_{\Sigma_j}\Bv\cdot\nv_j,
\end{eqnarray}
where ${[[}\phi_j{]]}=\phi_j|_{\Sigma_j^+}-\phi_j|_{\Sigma_j^-}=1$, from the construction of the basis functions \citep{alonso18}.  Thus, by enforcing zero magnetic flux through all the cutting surfaces of $\Omega$, the helicity in the domain can be written as in equation (\ref{hel_sc}). 

This approach to defining helicity in multiply connected domains has been used in several theoretical works \citep[e.g.][]{jordan98,faraco19}. For cases where there is non-zero magnetic flux, more work is needed to produce a gauge invariant helicity. We will now derive a more general formula which includes the zero flux condition as a special case. 

\subsection{Generalized Bevir-Gray formula}
Let us consider the following quantities,
\be
H_1=\int_{\Omega}\Av_1\cdot\Bv, \quad H_2 = \int_{\Omega}\Av_2\cdot\Bv,
\en
where $\Bv=\curl\Av_1$ and $\Bv=\curl\Av_2$. Without the zero flux condition, $H_1$ and $H_2$ are not, in general, gauge invariant in multiply connected domains. Considering the difference of these quantities,
\begin{eqnarray}
H_1-H_2 &=& \int_{\Omega}\Bv\cdot(\Av_1-\Av_2) = \int_{\Omega}\curl\Av_1\cdot(\Av_1-\Av_2) \nonumber \\
&=& \int_{\partial\Omega}\nv\times\Av_1\cdot(\Av_1-\Av_2) = \int_{\partial\Omega}\Av_1\times\nv\cdot\Av_2.
\end{eqnarray}
The second line follows from integration by parts and using $\curl(\Av_1-\Av_2)=\boldsymbol{0}$. Since $\curl\Av_1\cdot\nv=0=\curl\Av_2\cdot\nv$, the tangential traces $\Av_1\times\nv$ and $\Av_2\times\nv$ can be expressed, via a Helmholtz decomposition on $\partial\Omega$ \citep[e.g.][]{hiptmair12,alonso18}, as
\begin{eqnarray}
\Av_1\times\nv &=& \grad\eta\times\nv + \sum^g_{j=1}\alpha_j\brho_j\times\nv +  \sum^g_{j=1}\beta_j\brho'_j\times\nv, \label{expand1} \\ 
\Av_2\times\nv &=& \grad\xi\times\nv + \sum^g_{j=1}\delta_j\brho_j\times\nv +  \sum^g_{j=1}\mu_j\brho'_j\times\nv, \label{expand2}
\end{eqnarray} 
where $\eta$, $\xi$ are scalar functions and $\alpha_j, \beta_j, \delta_j, \mu_j\in\mathbb{R}$ $(j=1,\dots,g)$. \cite{alonso18} derived the following useful identities
\[
\int_{\partial\Omega}\grad\eta\times\nv\cdot\grad\xi = 0,
\]
\[
\int_{\partial\Omega}\grad\eta\times\nv\cdot\brho_j=0, \ \int_{\partial\Omega}\grad\eta\times\nv\cdot\brho'_i=0,
\]
\[
\int_{\partial\Omega}\brho_i\times\nv\cdot\brho_j = 0, \ \int_{\partial\Omega}\brho'_i\times\nv\cdot\brho'_j = 0,
\]
\[
\int_{\partial\Omega}\brho_j\times\nv\cdot\brho'_i = \delta_{ij} = -\int_{\partial\Omega}\brho'_i\times\nv\cdot\brho_j,
\]
for $1\le i, j\le g$. Note that $\delta_{ij}$ is the Kronecker delta. By making use of equations (\ref{expand1}) and (\ref{expand2}) and the above identities, it can be shown that
\be
\int_{\partial\Omega}\Av_1\times\nv\cdot\Av_2 = \sum_{j=1}^g\alpha_j\mu_j-\sum_{j=1}^g\beta_j\delta_j,
\en
where
\be
\alpha_j = \oint_{\gamma_j}\Av_1\cdot\tv_j, \ \beta_j = \oint_{\gamma'_j}\Av_1\cdot\tv'_j,
\en
\be
\delta_j = \oint_{\gamma_j}\Av_2\cdot\tv_j, \ \mu_j = \oint_{\gamma'_j}\Av_2\cdot\tv'_j.
\en
Since $\gamma_j'=\partial\Sigma_j$, we have, by Stokes' theorem,
\be
\beta_j = \oint_{\gamma'_j}\Av_1\cdot\tv'_j = \int_{\Sigma_j}\curl\Av_1\cdot\nv_j = \int_{\Sigma_j}\Bv\cdot\nv_j.
\en
By exactly the same reasoning,
\be
\mu_j = \oint_{\gamma'_j}\Av_2\cdot\tv'_j = \int_{\Sigma_j}\Bv\cdot\nv_j.
\en
Putting all these results together, we can now write
\be\label{blank}
H_1-H_2 =\int_{\Omega}\Av_1\cdot\Bv-\int_{\Omega}\Av_2\cdot\Bv= \sum_{j=1}^g\left(\oint_{\gamma_j}\Av_1\cdot\tv_j-\oint_{\gamma_j}\Av_2\cdot\tv_j\right)\left(\int_{\Sigma_j}\Bv\cdot\nv_j\right).
\en

An alternative route to equation (\ref{blank}) is to consider a result from \cite{blank57}. Based on the consideration of vector identities, they derive a particular formula (see their formula (6.5) on page 65) for the inner product of an irrotational vector $\xv$ and a solenoidal vector $\yv$,
\be
\int_{\Omega}\xv\cdot\yv = \sum_{j=1}^g\oint_{\gamma_j}\xv\cdot\tv_j\int_{\Sigma_j}\yv\cdot\nv_j,
\en
where the notation for cycles and surfaces is as before and the vector fields are everywhere tangent to the boundary. In our application, $\xv=\Av_1-\Av_2$ and $\yv=\Bv$.

Returning to equation (\ref{blank}), it is clear that the quantity
\be
\int_{\Omega}\Av\cdot\Bv -  \sum_{j=1}^g\left(\oint_{\gamma_j}\Av\cdot\tv_j\right)\left(\int_{\Sigma_j}\Bv\cdot\nv_j\right),
\en
is independent of the choice of vector potential. We are, therefore, led to define the gauge invariant magnetic helicity as
\be\label{hel_true}
{\Upsilon}(\Bv) = \int_{\Omega}\Av\cdot\Bv -  \sum_{j=1}^g\left(\oint_{\gamma_j}\Av\cdot\tv_j\right)\left(\int_{\Sigma_j}\Bv\cdot\nv_j\right).
\en
It is clear that for $g=1$, equation (\ref{hel_true}) reduces to the Bevir-Gray formula. Also, if $\Bv\perp\wh$ (zero flux), the helicity reduces to 
\be
{\Upsilon}(\Bv) = \int_{\Omega}\Av\cdot\Bv.
\en
This is also the case for simply connected domains, for which $\wh=\{\boldsymbol{0}\}$.



\subsection{Biot-Savart helicity}
As mentioned before, apart from the application of the Bevir-Gray formula in toroidal domains, helicity in multiply connected domains has, generally, taken the form of Biot-Savart helicity,
\be\label{hel_bs}
H(\Bv) = \int_{\Omega}\bs(\Bv)\cdot\Bv.
\en
The Biot-Savart helicity formula resembles that of helicity in simply connected domains, i.e. the second term on the right-hand side of equation (\ref{hel_true}) is missing.  We will now show that equations (\ref{hel_true}) and (\ref{hel_bs}) are  coincident.  Following \cite{cantarella01} and \cite{valli19}, it can be shown that the Biot-Savart operator, defined on $\Omega$, can be extended to $\mathbb{R}^3$.

For $\Bv\in\mathcal{V}$, $\bs(\Bv)\in\mathcal{V}$. Since $\Bv\cdot\nv=0$ on $\partial\Omega$ and $\div\Bv=0$ in $\Omega$, the extended magnetic field
\be
\widetilde{\Bv} = \left\{\begin{array}{cc}
\Bv & {\rm in}\ \Omega, \\
\boldsymbol{0} & {\rm in} \ \mathbb{R}^3\setminus\overline{\Omega},
\end{array}\right.
\en  
satisfies $\div\widetilde{\Bv}=0$ in $\mathbb{R}^3$. 

Equation (\ref{bs_op}) can be modified to
\be\label{bs_op_r}
\bs(\Bv)(\xv) = \frac{1}{4\pi}\int_{\mathbb{R}^3}\widetilde{\Bv}(\yv)\times\frac{\xv-\yv}{|\xv-\yv|^3}\,\d\yv.
\en
We know from \cite{cantarella01} that $\curl\bs(\Bv)=\widetilde{\Bv}$ and $\div{\Bv}=0$ in $\mathbb{R}^3$. Therefore, we can consider line integrals along closed paths and apply Stokes' theorem. 
\subsubsection{$n$-holed tori}
We know that each $\Sigma'_j\subset\Omega'$ $(j=1,\dots,g)$ is a surface bounded by a \emph{simple} cycle $\gamma_j$ (one with a connected boundary). We then have
\be\label{prop}
\oint_{\gamma_j}\bs(\Bv)\cdot\tv_j = \int_{\Sigma'_j}\curl\bs(\Bv)\cdot\nv'_j=0,
\en
since $\curl\bs(\Bv)=\boldsymbol{0}$ in $\Omega'$. Thus, choosing $\Av=\bs(\Bv)$ in equation (\ref{hel_true}), the helicity reduces to
\be\label{bs_hel}
\Upsilon(\Bv) =  \int_{\Omega}\bs(\Bv)\cdot\Bv.
\en
Note that equation (\ref{bs_hel}) can also be used for magnetic fields with zero flux, without having to impose this as a condition to ensure gauge invariance.

In above approach to deriving equation (\ref{bs_hel}) we have made the standard construction of extending the Biot-Savart operator to $\mathbb{R}^3$ by assuming that the magnetic field is zero outside the domain.  {By virtue of equation (\ref{bs_hel}),} magnetic field in the domain can inherit the field line topology interpretation of \cite{moffatt69} and \cite{arnold92}. For applications where linkage with magnetic field outside the domain is important, {see Section \ref{sec_mutual} below.} {First, however, we will investigate how the Biot-Savart operator can be used to understand the linkage of field lines on the surface of the domain with the domain itself.}


\subsubsection{Field line helicity on a toroidal boundary}
Although the {second term on the right-hand side of equation (\ref{hel_true}), related to the domain topology,} is zero for Biot-Savart helicity, this does not mean that integrals on the domain boundary are unimportant. For the case of a standard torus $(g=1)$, the property in (\ref{prop}) can be extended to define the \emph{field line helicity} \citep[e.g.][]{berger88,yeates13} on the boundary when the path of integration follows a closed field line. Such field lines are possible everywhere on the boundary of a torus as the Euler characteristic of the boundary is zero. A prominent example of such a field line is a \emph{torus knot} \citep[e.g.][]{oberti18}. For domains with $g>1$, closed field lines on the boundary are possible but not everywhere on the boundary. {This is  due to the Euler characteristic being non-zero and, as a consequence of the `hairy ball' theorem \citep[e.g.][]{frankel04}, smooth vector fields on the surfaces of these domains must have at least one point where they vanish}.

{
A closed path $\gamma$ on the surface of a
torus $\Omega$ can be expressed in terms of the basis cycles $\gamma_1$ and $\gamma_1'$ as
\be
\gamma = L' \gamma_1 + L \gamma'_1,
\en
up to bounding cycles on $\partial \Omega$, that are homologically trivial. Here $L' \in \mathbb Z$ is the linking number between $\gamma$ and  $\gamma'_1$ (slightly deformed outside $\Omega$, in order that there is no intersection with $\gamma$); similarly, $L\in\mathbb Z$ is the linking number between $\gamma$ and  $\gamma_1$ (slightly deformed inside $\Omega$, in order that there is no intersection with $\gamma$). Therefore, if $\wv$ is a vector field such that $\curl\wv \cdot \nv = 0$ on $\partial \Omega$ and $\tv$ is the unit tangent vector on $\gamma$, 
\be
\oint_{\gamma} \wv\cdot \tv =  L' \oint_{\gamma_1} \wv\cdot \tv_1 +  L \oint_{\gamma_1'} \wv\cdot \tv'_1.
\en
Applying this result to the Biot-Savart vector field $\bs(\Bv)$, for which (from (\ref{prop})),
\be
\oint_{\gamma_1} \bs(\Bv) \cdot {\tv}_1 = 0,
\en
it follows that
\be
\oint_{\gamma} \bs(\Bv)\cdot {\tv} =  L \oint_{\gamma_1'} \bs({\Bv})\cdot {\tv}_1'.
\en
Thus, by Stokes' theorem,
\be\label{fl1}
\oint_{\gamma} \bs({\Bv})\cdot {\tv} = L \int_{\Sigma_1} {\Bv} \cdot {\nv},
\en
where $\Sigma_1$ is a cutting surface of $\Omega$ with $\partial \Sigma_1 = \gamma'_1$. The value $L $ can be  interpreted as the number of loops of $\gamma$ around the minor cross section of the torus (say, the `linking number' of the field line with the torus). 

It is interesting to note that the field line helicity (the integral on the left-hand side of equation (\ref{fl1})) `knows' about the linkage of the boundary curve with field lines inside the torus, by virtue of equation (\ref{bs_op_r}). This result is true for any magnetic field inside the torus, no matter how complex the field line topology. The value of the field line helicity, however, depends simply on the magnetic flux and a purely topological quantity depending only on the curve and the domain. 
}
%
{Equation (\ref{fl1}) is analogous to the voltage formula of a transformer (for a topological perspective on this formula, see \cite{gross04}).}

\subsubsection{Field line helicity on the boundary of a toroidal shell}
The Biot-Savart helicity is coincident with the general helicity, equation (\ref{hel_true}), in a toroidal shell by the same procedure as described for $n$-tori. The toroidal shell does, however, have some extra physical interpretations related to the cutting surfaces and field line helicities on the boundaries.

As described in Section \ref{sec:geo_setup}, some care is required in order to identify the $\Sigma_j$ cutting surfaces. Given suitable cutting surfaces, the helicity for a toroidal shell is
\be\label{tt}
\Upsilon(\Bv) = \int_{\Omega}\Av\cdot\Bv - \oint_{\gamma_1}\Av\cdot\tv_1\int_{\Sigma_1}\Bv\cdot\nv_1-\oint_{\gamma_2}\Av\cdot\tv_2\int_{\Sigma_2}\Bv\cdot\nv_2,
\en 
where the notation is standard. The cutting surface $\Sigma_1$ is that shown in Figure \ref{major} and $\Sigma_2$ is that shown in Figure \ref{minor}. Therefore, we can label the fluxes as
\be\label{fluxes}
\int_{\Sigma_1}\Bv\cdot\nv_1 = \Psi_P, \ \int_{\Sigma_2}\Bv\cdot\nv_2 = \Psi_T,
\en
where $\Psi_P$ and $\Psi_T$ are the poloidal and toroidal fluxes respectively. If we select $\Av=\bs(\Bv)$, then the general helicity formula naturally reduces to the form of equation (\ref{bs_hel}) due to the property in (\ref{prop}).

Considering the field line helicity in (\ref{fl1}), there are now two separate boundaries for a (closed) field line to lie on. For a field line lying on the outer boundary of a toroidal shell, the interpretation is the same as that of a standard torus. That is, a field line twisting around the minor circumference of the torus $L$ times will have a field line helicity of $L\Psi_T$. For a field line on the inner boundary, the path of the line integral can also be deformed into a union of circles around the major and minor circumferences of the toroidal hole. This time, however, due to the property in (\ref{prop}), the contribution to the integral from circles around the minor cross section of the hole is zero. The non-zero contribution comes from the major cross section and if a field line on the inner boundary wraps around the major cross section $L$ times, its field line helicity is $L\Psi_P$.

{
\subsection{Mutual helicity}\label{sec_mutual}
Through the summation term on the right-hand side of equation (\ref{hel_true}), the domain topology enters explicitly into the helicity formula for multiply connected domains. This term is related to the \emph{mutual helicity} \citep{laurence93,cantarella00b}, a quantity that measures the linkage of the magnetic field in two domains, $\Omega$ and $\Omega'$ say. 
Indeed, through equation (\ref{hel_true}) we can prove a representation formula for the mutual helicity which is equivalent to that obtained by \cite{cantarella00b} for domains $\Omega$, $\Omega'\subset\mathbb{R}^3$ of arbitrary topology.

With its topological connection to the Gauss linking number \citep[e.g.][]{moffatt69,laurence93}, helicity is often written as 
\be
H_U(\Bv) = \frac{1}{4\pi}\int_{U\times U}\left(\Bv(\xv)\times\Bv(\yv)\cdot\frac{\xv-\yv}{|\xv-\yv|^3}\right)\,{\rm d}\xv\,{\rm d}\yv,
\en
where $U$ is a bounded open set (but not necessarily connected) and $\Bv$ is a magnetic field tangent to $\partial U$. If we consider two disjoint bounded domains $M$ and $N$ and we set $U=M\cup N$ and $\Bv_M = \Bv_{|M}$, $\Bv_N = \Bv_{|N}$, then it is easily checked that
\be\label{hel_exp}
H_{M\cup N}(\Bv) = H_M(\Bv_M)+H_N(\Bv_N) + 2H(\Bv_M,\Bv_N),
\en
where
\be
H(\Bv_M,\Bv_N) = \frac{1}{4\pi}\int_{M\times N}\left(\Bv_M(\xv)\times\Bv_N(\yv)\cdot\frac{\xv-\yv}{|\xv-\yv|^3}\right)\,{\rm d}\xv\,{\rm d}\yv
\en
is the mutual helicity. The \emph{self helicities}, $H_M(\Bv_M)$ and $H_N(\Bv_N)$, can be readily expressed by means of the Biot-Savart operator,
\be
H_M(\Bv_M) = \int_M\bs(\Bv_M)\cdot\Bv_M, \quad H_N(\Bv_N) = \int_N\bs(\Bv_N)\cdot\Bv_N.
\en
Consider an open cube $Q$ such that $\overline{M\cup N}\subset Q$. In this domain, we extend $\Bv$ by $\boldsymbol{0}$ in $Q\setminus\overline{M\cup N}$. We label this extension $\widehat{\Bv}$ and note that it is divergence-free in $Q$ and tangent to $\partial Q$. Further, its helicity in $Q$ coincides with that of $\Bv$ in $M\cup N$,
\be
H_Q(\widehat{\Bv}) = H_{M\cup N}(\Bv).
\en
Since $Q$ is simply connected, we can use equation (\ref{hel_true}) with $g=0$ to write
\be
H_Q(\widehat{\Bv}) = \int_Q\widehat{\Av}\cdot\widehat{\Bv},
\en
where $\widehat{\Av}$ is any vector potential of $\widehat{\Bv}$ in $Q$. Since $\widehat{\Bv}$ is vanishing outside $M\cup N$, it follows that
\be
H_Q(\widehat{\Bv}) = \int_M\widehat{\Av}_{|M}\cdot\widehat{\Bv}_M + \int_N\widehat{\Av}_{|N}\cdot\widehat{\Bv}_N.
\en
Using equation (\ref{hel_true}) in $M$ (with genus $g_M$) and in $N$ (with genus $g_N$) we have
\be\label{gg1}
\int_M\widehat{\Av}_{|M}\cdot\widehat{\Bv}_M = \Upsilon_M(\Bv_M) + \sum_{j=1}^{g_M}\left(\oint_{\gamma_j^M}\widehat{\Av}_{|M}\cdot\tv_j^M\right)\left(\int_{\Sigma_j^M}\Bv_M\cdot\nv_j^M\right)
\en
and
\be\label{gg2}
\int_N\widehat{\Av}_{|N}\cdot\widehat{\Bv}_N = \Upsilon_N(\Bv_N) + \sum_{l=1}^{g_N}\left(\oint_{\gamma_l^N}\widehat{\Av}_{|N}\cdot\tv_l^N\right)\left(\int_{\Sigma_l^N}\Bv_N\cdot\nv_l^N\right).
\en
Using equation (\ref{bs_hel}), $H_M(\Bv_M) = \Upsilon(\Bv_M)$ and $H_N(\Bv_N) = \Upsilon(\Bv_N)$. Substituting these results into equations (\ref{gg1}) and (\ref{gg2}) and then substituting these into equation (\ref{hel_exp}), the mutual helicity is found to be
\begin{eqnarray}
H(\Bv_M,\Bv_N) &=& \frac12\left[\sum_{j=1}^{g_M}\left(\oint_{\gamma_j^M}\widehat{\Av}_{|M}\cdot\tv_j^M\right)\left(\int_{\Sigma_j^M}\Bv_M\cdot\nv_j^M\right)\right. \nonumber\\ 
&& +\left.\sum_{l=1}^{g_N}\left(\oint_{\gamma_l^N}\widehat{\Av}_{|N}\cdot\tv_l^N\right)\left(\int_{\Sigma_l^N}\Bv_N\cdot\nv_l^N\right)\right].
\end{eqnarray}
A careful analysis of the values of the linking numbers between $\gamma^M_j$ and $\gamma^N_l$ would show that this formula is equivalent to that given in \cite{cantarella00b}. For simplicity, let us show that this is true in the case of two linked solid tori \citep{moffatt69,laurence93,cantarella00b}. We have $g_M=1$ and $g_N=1$; the cycle $\gamma^M_1$ is the boundary of a surface $\Sigma'_M$, contained in $Q\setminus \overline{M}$, and the cycle $\gamma^N_1$ is the boundary of a surface $\Sigma'_N$, contained in $Q\setminus \overline{N}$. Let us also assume that the unit tangent vector $\tv^M_1$ crosses $\Sigma_M$, the cross section of the torus $M$, consistently with the direction of normal vector $\nv_{\Sigma_M}$ on it, and similarly for $\tv^N_1$ and the cross section of $N$. In this way the linking number of $M$ and $N$ has value 1. By Stokes' theorem,
\be
\oint_{\gamma_1^M}\widehat{\Av}_{|M}\cdot\tv_1^M = \int_{\Sigma'_M}\curl\widehat{\Av}\cdot\nv_{\Sigma'_M} = \int_{\Sigma'_M}\widehat{\Bv}\cdot\nv_{\Sigma'_M}.
\en
Since $\widehat{\Bv}=\boldsymbol{0}$ outside $M\cup N$, it follows that
\be
\int_{\Sigma'_M}\widehat{\Bv}\cdot\nv_{\Sigma'_M} = \int_{\Sigma'_M\cap N}\Bv_N\cdot\nv_{\Sigma'_M} = \int_{\Sigma_N}\Bv_N\cdot\nv_{\Sigma_N}.
\en
Similarly,
\be
\oint_{\gamma_1^N}\widehat{\Av}_{|N}\cdot\tv_1^N = \int_{\Sigma_M}\Bv_M\cdot\nv_{\Sigma_M},
\en
and we obtain
\be
H(\Bv_M,\Bv_N) = \left(\int_{\Sigma_M}\Bv_M\cdot\nv_{\Sigma_M}\right)\left(\int_{\Sigma_N}\Bv_N\cdot\nv_{\Sigma_N}\right).
\en

}

\section{Summary}
In this paper, we have unified the main approaches to calculating magnetic helicity in multiply connected domains. The correct approach for determining a gauge invariant helicity is to consider the full Helmholtz decomposition, rather than the standard gauge transformation suitable for simply connected domains. We derive a gauge invariant expression for helicity that generalizes the Bevir-Gray formula for a torus and is suitable for any connected, bounded domain in $\mathbb{R}^3$, no matter how complicated the topology.

The discovery of magnetic helicity as a topological quantity was originally made by examining Biot-Savart helicity \citep{moffatt69}. We show that the general helicity {formula, which holds for any vector potential, can naturally} reduce to Biot-Savart helicity.  The line integrals of the Biot-Savart operator on closed paths on the domain boundary can be interpreted as field line helicities and these arise naturally from the general formula (\ref{hel_true}). {The general form for mutual helicity is also shown to follow directly from equation (\ref{hel_true}).}  

In several works, a gauge invariant helicity is found by imposing the constraint of zero flux outside the domain in question, e.g. \cite{chui95} for knotted domains and \cite{taylor15,hussain17} for periodic toroidal domains. The helicities in these works are coincident with Biot-Savart helicity by virtue of the property in (\ref{prop}).


What is clear from the general helicity formula is that the topology of the domain must be taken into account for a correct interpretation of quantities in the domain. Multiply connected domains are used widely in the context of MHD simulations. For example, a cube with two identified boundaries is equivalent to a torus topologically. Similarly, a cube with two pairs of identified boundaries is equivalent to a toroidal shell topologically. If, in such periodic domains, the magnetic field is tangent to the non-periodic boundaries, then the helicity can be interpreted in terms of equation (\ref{hel_true}) and, by extension, the Biot-Savart helicity (\ref{hel_bs}). Care must be taken, however, when interpreting helicity (and other quantities) for evolving magnetic fields in periodic domains. For example,  \cite{berger97} discusses how periodic domains lead to strange results, such as magnetic flux ropes turning themselves inside out through magnetic reconnection. The issue here is one of mathematical modelling. That is, if a periodic domain is used to model a magnetic field in a simply connected domain, then there may be some unwanted effects due to the topology of the domain.  For the (2D) flux tube example in \cite{berger97} (see Figure 3 of that work), the behaviour shown makes perfect sense when considering the topology of the domain, the surface of a torus in this case. The behaviour, however, is physically unrealistic for a simply connected domain. 

Another periodic domain that is popular in MHD simulations is the triply periodic cube. This domain, however, cannot be be embedded in $\mathbb{R}^3$ and equation (\ref{hel_true}) does not apply in this case. Although it is possible to develop topological invariants in this domain \citep[e.g.][]{deturck13}, the physical significance of these quantities remains to investigated in depth.  



\end{document}